\documentclass[aps,prb,onecolumn,,floatfix]{revtex4-2}

\usepackage{amsmath}
\usepackage{amssymb}
\usepackage{graphicx, float}

\newcommand{\bea}{\begin{eqnarray}}
\newcommand{\eea}{\end{eqnarray}}

\newcommand{\parent}[1]{\left( #1 \right)}

\begin{document}

\title{Making sense of nonequilibrium current fluctuations: A molecular motor example}

\author{David Andrieux}

\noaffiliation

\begin{abstract}
The nonequilibrium response and fluctuations of Markovian systems, both near and far from equilibrium, are best understood by varying their parameters along dynamical equivalence classes. 
In this note, I illustrate this approach for an analytically solvable molecular motor toy model.
\end{abstract}


\maketitle

My latest research has been focused on understanding how the dynamical features of a system shape its transport and thermodynamic properties.
This has led to a series of new insights on the nonequilibrium current response and fluctuations. 
For example:
\begin{itemize}

\item Equivalence classes exist in the space of stochastic dynamics that determine their (large) fluctuations \cite{A12b}

\item As a consequence, equilibrium and nonequilibrium fluctuations are identical \cite{A12c}

\item The currents fluctuations can be obtained from the current response curve itself, and both can be made fully symmetric. The response coefficients are then expressed in terms of equilibrium fluctuations. \cite{A22}

\item The current fluctuations can be obtained by solving a system of polynomial equations instead of the traditional eigenvalue analysis \cite{A23}

\end{itemize}

The problem is that these results are scattered over multiple papers, sometimes with multi-year gaps between them (I do physics research on the side, and only when I have both the time and the interest, so it's been on and off - well, mostly off - over the last years). 
This makes it hard to see the logical thread between these results and that, taken together, they offer a new way to look at nonequilibrium systems.
Also, I prefer 'short-and-to-the-point' articles when presenting my findings, which has the benefits of minimizing my workload but can make it harder for other researchers to access my ideas.


To address these points I could write a review article bringing everything together (and maybe I will at some point). 
But in the short term, I think it's more informative to illustrate these various results on a simple model that can be solved analytically. 

\section{A molecular motor toy model}

I consider a Markov chain representing a molecular motor with $2L=6$ states corresponding to different conformations of the protein complex. 
These states form a cycle of periodicity $2L$ corresponding to the revolution by 360° for a rotary motor or the reinitialization for a linear motor.
The motor alternates between two types of states according to the transition matrix
\bea
P&=&
\begin{pmatrix}
0 &  p_1 &  &  &  & 1-p_1\\
1-p_2 &  & p_2 &  &  & \\
 &  1-p_1 &  & p_1 & & \\
 &   & \ddots & 0 & \ddots & \\
 &   &  & 1-p_1 & 0 & p_1\\
p_2  &  &  &  & 1-p_2 & 0
\end{pmatrix}_{2L \times 2L} 
\label{toymodel}
\eea
This is a discrete-time version of a model that I studied together with Pierre Gaspard \cite{AG06}; a continuous-time and -space model was further developed by Pierre Gaspard and Eric Gerritsma \cite{GG07}.

The matrix $P$ is doubly stochastic, so that its stationary state is given by the uniform distribution ${\bf \pi}= (1,1,1,1,1,1)/2L$ for all parameters $(p_1,p_2)$.
The average current $J$ and affinity $A$ take the form
\bea
J = \frac{1}{2L} \parent{p_1+p_2 -1}
\label{J}
\eea
and
\bea
A = L \, \ln \frac{p_1p_2}{(1-p_1)(1-p_2)} \, .
\label{A}
\eea
The system is at thermodynamic equilibrium when $J = A = 0$. 
This corresponds to the line $(p_0, 1-p_0)$ for $p_0 \in [0,1]$ in the space of parameters $(p_1,p_2)$.
More generally, the time-reversed dynamics of $(p_1,p_2)$ is given by $(1-p_2,1-p_1)$.

\section{This looks simple enough (until it isn't)}

\begin{figure}[h]
\includegraphics[scale=.5]{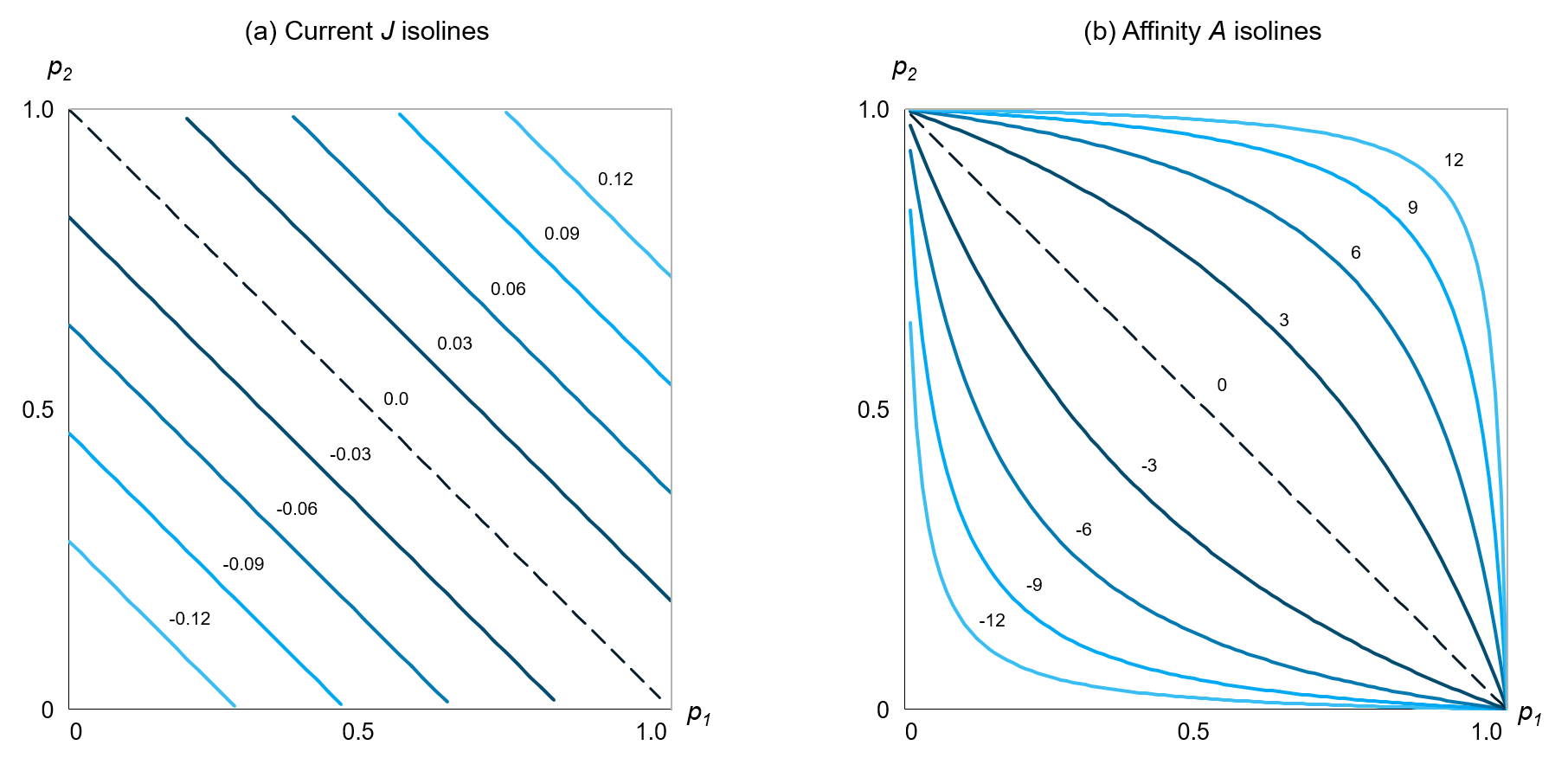}%
\caption{{\bf Isolines for the current and affinity in the parameter space}. (a) The current isolines are equally spaced straight lines. 
(b) Isolines for the affinity $A$. In both figures the dashed line correponds to the thermodynamic equilibrium line.
From these two plots we see that no unique relationship exists between currents and affinities.}
\label{fig1}
\end{figure}

It is worth visualizing how the current (\ref{J}) and affinity (\ref{A}) vary across the parameter space $(p_1,p_2)$. 
The current $J$ is a linear combination of $(p_1,p_2)$, which gives straight isolines (i.e. the curves for which the values of $J$ is constant). 
The affinity $A$ varies non-linearly with the parameters $(p_1,p_2)$, giving rise to curved isolines that get denser further away from equilibrium (Fig. \ref{fig1}).

\begin{figure}[h]
\includegraphics[scale=.62]{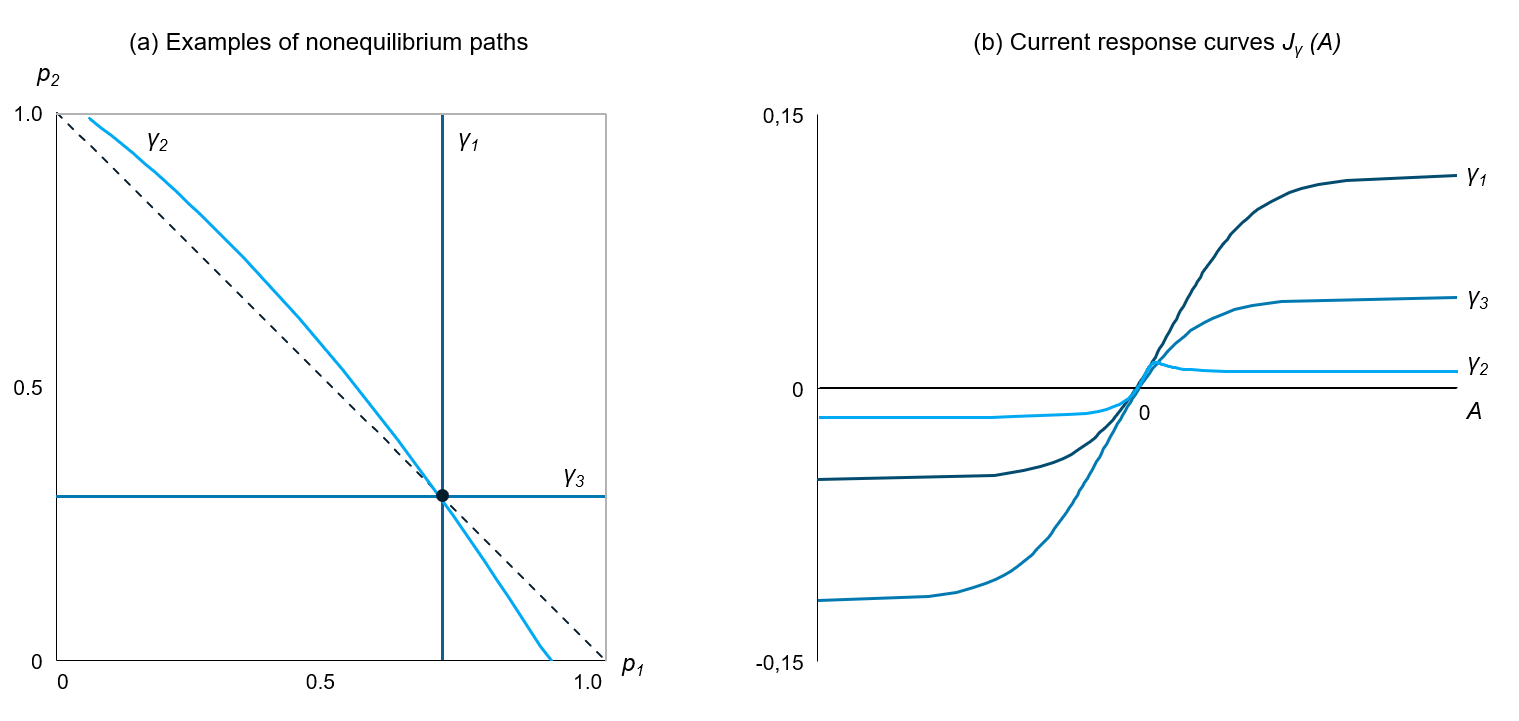}%
\caption{{\bf Current response for different paths in the parameter space}. (a) Examples of nonequilibrium paths $\gamma$ going through the equilibrium state $(p_0, 1-p_0)=(0.7, 0.3)$. 
(b) Each path $\gamma$ determines a different current response curve (affinities range from $\pm 25$). 
The linear response is identical across all paths $\gamma$ (for a given $p_0$), but the nonlinear response is different. 
The response curves do not show any particular symmetries and reach different plateaus. The response curve $\gamma_2$ is even non-monotomic.}
\label{fig2}
\end{figure}

From these two graphs we can already make an important observation: the current response $J(A)$ is not uniquely defined.
Indeed, the current response curve will look different depending on the path $\gamma$ followed in the parameter space (Fig. \ref{fig2}). 

So what most people do (myself included before developping the present approach) is to fix one of the variables, say $p_1$, and express $p_2$ as a function $g(p_1,A)$. 
This corresponds to the path $\gamma_1$ in Figure \ref{fig2}, while the path $\gamma_3$ corresponds to fixing $p_2$ instead. 
However, this affects the shape of the currents and other transport properties in an uncontrolled and not necessarily representative way (we'll see a striking example in the next section when looking at the diffusion coefficient).

Even tough this observation is in many ways straightforward, it is not always properly appreciated in the litterature (and this has confused me as well in the past). 
One reason is that the {\it linear} response of a system is independent of the path $\gamma$. This is, however, {\it no longer true in the nonlinear regime} \cite{H05}.
For example, in the toy model (\ref{toymodel}) 
\bea
J^\gamma (A) = \frac{p_0(1-p_0)}{2 L^2} A + f(\gamma) A^2 + ...
\label{LR}
\eea
That is, the linear response indeed depends on the equilibrium state $(p_0,1-p_0)$ only, and not on the path $\gamma$ in the parameter space.
However, higher-order coefficients and the overall response curve do depend on $\gamma$, and the (arbitrary) choice of path $\gamma$ essentially determines the shape of the response curve (Fig. \ref{fig2}).  

\section{Making sense of the nonlinear regime}

At this point you might be wondering: Is affinity even a helpful concept in the nonlinear regime? 
And, if everything is path dependent, can we still obtain a meaningful description of nonequilibrium thermodynamics?
What is even the 'right' equilibrium state to consider?

An important result to start addressing these questions is the {\it fluctuation theorem}.
The fluctuation theorem restricts the shape of the current fluctuations, and is valid arbitrarily far away from equilibrium. 
It can thus be used to derive relationships between the nonlinear response of a system and its fluctuations \cite{AG04}. 

That said, how instructive really is the fluctutation theorem? 
One the one hand, it holds across many classes of systems. 
On the other hand, it merely reflects the effect of time reversal on a system. 
So while its implications are broad, they are also, by construction, generic. 

But we can go one step further than the fluctuation theorem thanks to new relations that (1) reflect the dynamics of the specific system under study, and (2) are still applicable to any nonequilibrium system.
This is what I want to illustrate in the rest of this note. 

To this end let's partition the parameter space $(p_1,p_2)$ into equivalence classes \cite{A12c}.
These equivalence classes provide the missing ingredient to study transport in the nonlinear regime.
They can be defined for any system, but their shape is specific to each individual system. 

These dynamical equivalence classes are determined by an equilibrium dynamics modulated by all possible affinities \cite{A22}.
For the toy model (\ref{toymodel}), the equivalence classes can thus be parametrized by the equilibrium points $(p_0, 1-p_0) = (\alpha, 1-\alpha)$.
Then, a point $(p_1,p_2)$ belongs to the equivalence class $[\alpha]$ if
\bea
\alpha = \frac{\sqrt{ p_1 (1-p_2) } }{\sqrt{ p_1 (1-p_2) } + \sqrt{ p_2 (1-p_1) } } \, . 
\label{alpha}
\eea
Note that if a dynamics $(p_1,p_2)$ belongs to $[\alpha]$ then its time-reversal $(1-p_2,1-p_1)$ also belongs to $[\alpha]$.

Equivalently, introducing $\kappa_\alpha = \alpha/(1-\alpha) = p_0/(1-p_0)$, the classes $[\alpha]$ can be parametrized by the equilibrium state $(\alpha,1-\alpha)$ and the affinity $A$ as
\bea
p^\alpha_{1}(A) = \frac{\kappa_\alpha \exp(A/2L)}{1+ \kappa_\alpha \exp(A/2L)}, \quad \quad p^\alpha_{2}(A)  = \frac{\kappa^{-1}_\alpha \exp(A/2L)}{1+ \kappa^{-1}_\alpha \exp(A/2L)} \, .
\label{param}
\eea

\begin{figure}[h]
\includegraphics[scale=.62]{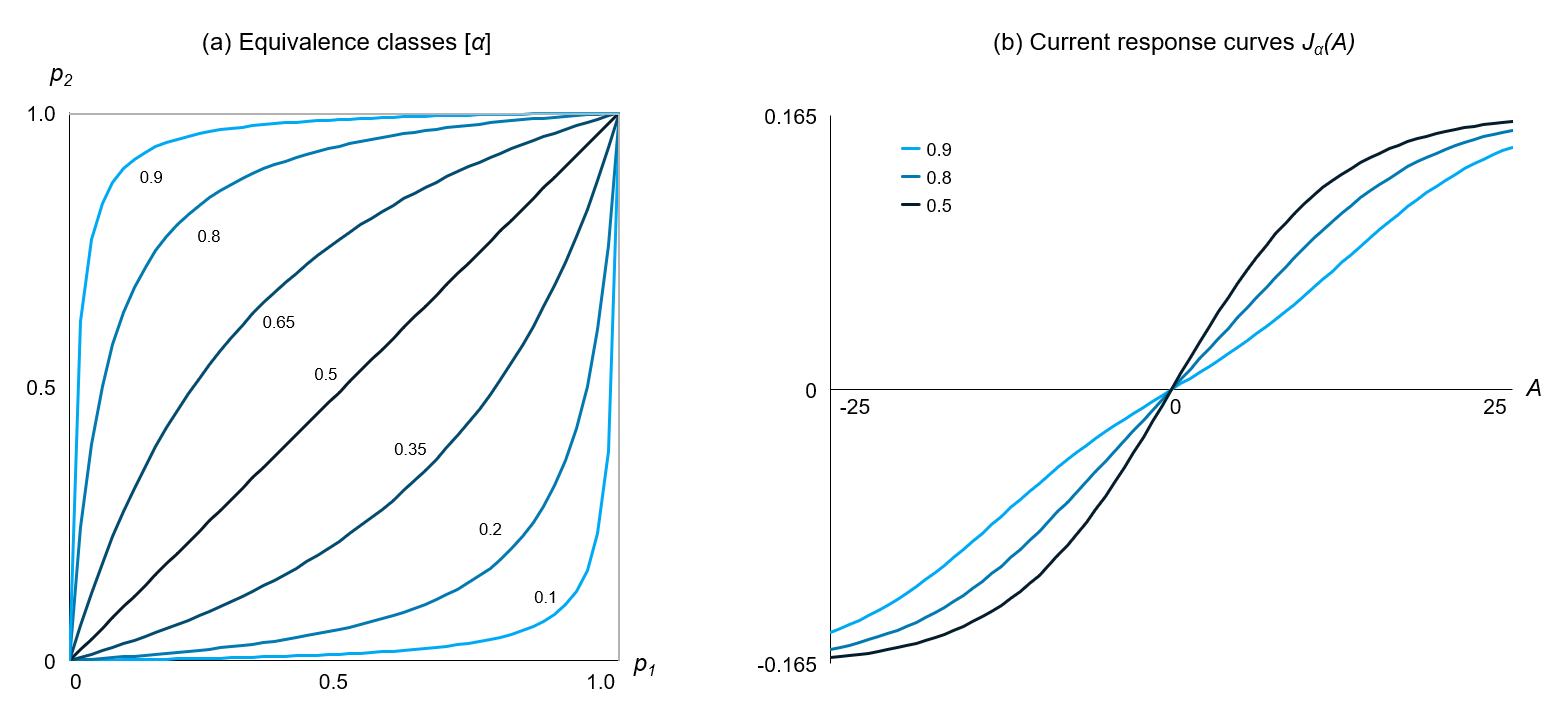}%
\caption{{\bf Dynamical equivalence classes $[\alpha]$ in the parameter space and their corresponding current responses}. (a) Equivalence classes $[\alpha]$ determined by Eq. (\ref{alpha}). Each class crosses the equilibrium once at $p_0 = \alpha$ and spans the full range of affinities.
(b) Corresponding current responses $J^\alpha$. All current response curves are monotonic, antisymmetric, and reach the minimal and maximal possible currents $J = \pm (1/2L)$ as $A \rightarrow \pm \infty$.}
\label{fig3}
\end{figure}

Each equivalence class determines a curve in the space $(p_1,p_2)$ that goes through an equilibrium dynamics and crosses each current and affinity isolines only once (Fig. \ref{fig3}a).
We thus visually verify that, when varying parameters along one of the equivalence class, the current and affinity are uniquely defined.
In addition, this leads to a current response function that is antisymmetric and reaches the minimal and maximal possible currents (Fig. \ref{fig3}b). 
The symmetry of $J^\alpha$ arises directly from the fact that both a dynamics and its time-reversal belongs to the same equivalence class.\\

So why are these equivalence classes special? 
After all, the space of parameters can be partitioned in many ways. 
What is interesting about this decomposition is that it emerges from the {\it intrinsic dynamics} of the system (specifically from the behavior of the matrix $(P\circ P^T)^{(1/2)}$ where $P^{(1/2)}$ denotes the elementwise exponentiation, see Appendix for details). 
And, even more importantly, it brings new properties that are only valid when using this decomposition. 
This is what I will demonstrate next.

\section{I'm intrigued. What other perks do I get?}

In the toy model (\ref{toymodel}) the mechanical and chemical affinities are tightly coupled so that we only have one independent current. 
Unfortunately this means I cannot illustrate the symmetries that appear when multiple independent currents and affinities are present \cite{A22}.
What I can do though is examine the current fluctuations. 

The current fluctuations are characterized by their cumulants $K_n (A)$. 
Together, they encode all the information about the statistical properties of the current: $K_1 = J$ is the mean current, $K_2$ the variance or diffusion coefficient, $K_3$ the skewness and so on (note that diffusion is a term that comes from physics while skewness comes from the field of probability - that's because, until recently, higher-order statistics were too difficult to observe and thus they didn't get their own physics names).

\begin{figure}[h]
\includegraphics[scale=.62]{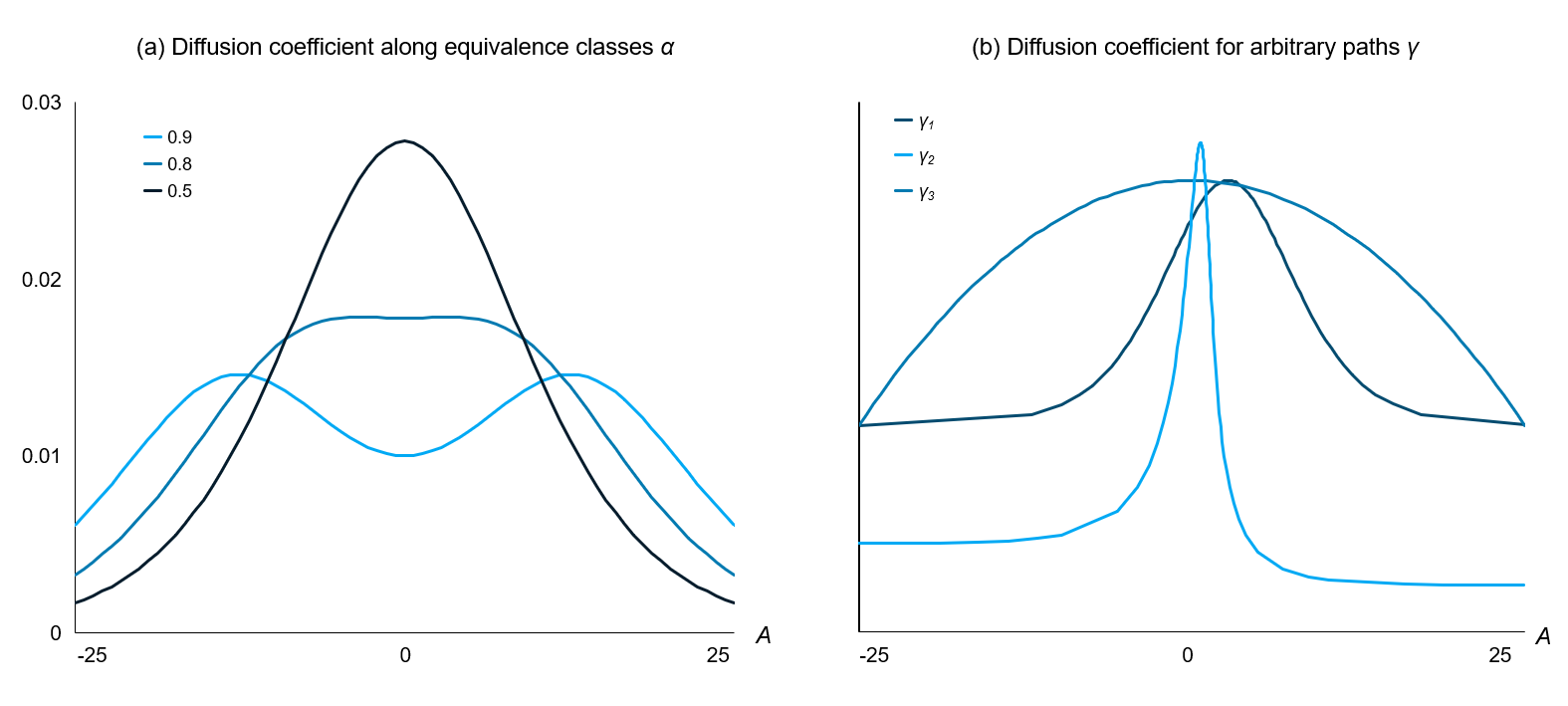}%
\caption{{\bf Diffusion coefficient for the current fluctuations}. 
(a) The diffusion coefficients $K^\alpha_2$ are symmetric with respect to $A$ within an equivalence class and vanish as $A \rightarrow \pm \infty$. 
They are obtained by deriving the currents $J^\alpha$ according to Eq. (\ref{K}).
(b) In contrast, for arbitrary paths $\gamma$ the diffusion coefficients take complex, non-symmetric shapes reflecting the arbitrary nature of the paths in parameter space. 
As a result, no direct connection exists between the diffusion coefficients $K^\gamma_2$ and the currents $J^\gamma$.
The paths $\gamma$ shown here are the same as in Figure \ref{fig2}.}
\label{fig4}
\end{figure}

Remarkably, within an equivalence class, we can {\it obtain all the cumulants directly from the current} \cite{A22}:
\bea
K_{n+1}^{\alpha} (A) = 2^{n} \frac{{\rm d}^n J^\alpha (A)}{{\rm d} A^n}     \, .
\label{K}
\eea
In other words, knowing the current $J^\alpha (A)$ along an equivalence class $[\alpha]$ determines the entire behavior of the current fluctuations, both close and far from equilibrium! 
Therefore, the higher-order statistics can all (litteraly) be derived from the current $J^\alpha(A)$ without any additional calculation or measurement.
This relation is illustrated in Figure (\ref{fig4}a) for the diffusion coefficient $K^\alpha_2 (A)$.
In fact, using Eqs (\ref{param}) and that $p'_{i} = p^\alpha_{i} (1-p^\alpha_{i})/(2L)$ for $i=1,2$, the diffusion coefficient takes the form
\bea
K_2 (A) = \frac{2}{(2L)^2} \Big[p_1(1-p_1) + p_2(1-p_2) \Big] \, .
\label{K2}
\eea
This formula actually holds for any $(p_1,p_2)$. Higher-order cumulants can be derived by further derivation.
We also recover the linear response (\ref{LR}) according to $dJ/dA(0) = K_2(0)/2$. 
More generally, Eq. (\ref{K}) reveals that {\it all current response coefficients can be expressed in terms of equilibrium fluctuations} \cite{A22}.

Since the current is antisymmetric along an equivalence class, $J^\alpha (A) = - J^\alpha (-A)$, the diffusion coefficient (\ref{K2}) is symmetric in $A$. 
Also, it vanishes in the limit of large positive and negative affinities. 
This results from the shape of the equivalence classes, which all start at $(p_1,p_2) = (0,0)$ and end at $(p_1,p_2) = (1,1)$ where the dynamics is deterministic.

This contrats with the behavior for paths $\gamma$ outside the equivalence classes, where the diffusion coefficients take complex, non-symmetric shapes (Fig. \ref{fig4}b) due to the somewhat arbitrary nature of these paths in the parameter space. 
More importantly, this also means that there is no direct connection similar to Eq. (\ref{K}) between the diffusion coefficient and the current, and that the diffusion coefficient needs to be evaluated independently. 

Similar observations hold for the higher-order cumulants. 
All even cumulants are either symmetric or anti-symmetric, $K^\alpha_n (A) = (-1)^n K^\alpha_n (-A)$, depending on whether $n$ is even or odd.
They all vanish in the limit of infinite positive or negative affinities.
They can directly by obtained from the currents from Eq. (\ref{K}), while the cumulants for arbitrary paths won't have any direct connection with the corresponding currents and will need to be calculated or measured independently. 
Again, the relation (\ref{K}) does not hold outside an equivalence class, showing that these results go beyond the traditional fluctuation theorem.\\

Now, when studying current fluctuations the typical starting point is the so-called generating function, which also encodes the same information as the cumulants \cite{AG04}. 
As clear from my exposition, the generating function is no needed ${\it per se}$ as all the information is encoded in the current response. 
Nonetheless, it still provides a useful theoretical and practical formulation of the problem.
So for completeness let me show how to obtain the generating function within this framework. 

\begin{figure}[h]
\includegraphics[scale=.7]{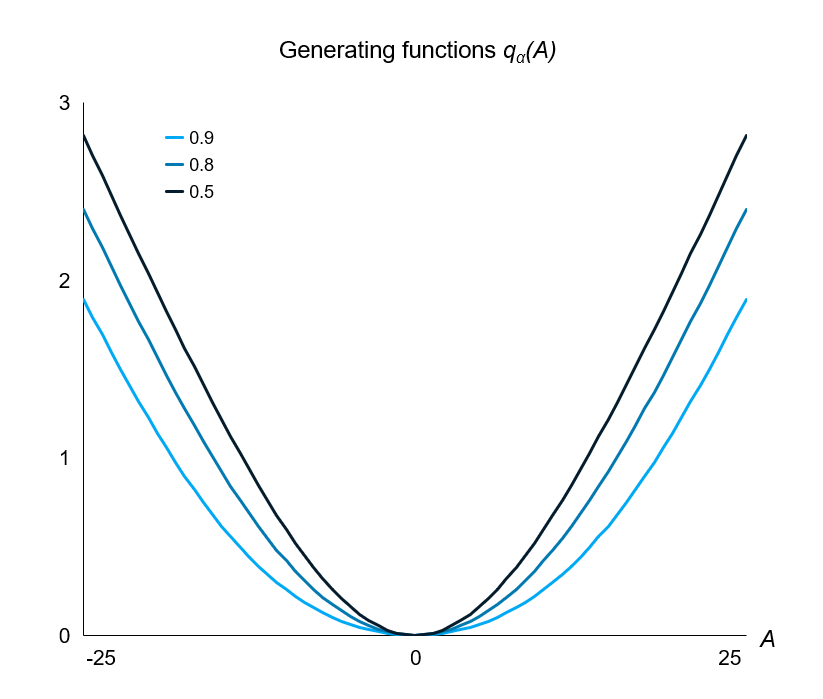}%
\caption{{\bf Current generating functions for different equivalence classes $\alpha$}. The generating functions (\ref{q}) are centered around equilibrium and are symmetric with respect to $A$. The generating function of a nonequilibrium dynamics $(p_1,p_2) \in [\alpha]$ is obtained by a translation of $q^\alpha$. 
The currents $J^\alpha (A)$ are obtained by deriving $q^\alpha (A)$ (not shown).
}
\label{fig5}
\end{figure}

The key quantity to consider is the scalar field \cite{A12b}
\bea
\rho = \sqrt{p_1 (1-p_2)} + \sqrt{p_2 (1-p_1)} \, ,
\label{rho}
\eea
which corresponds to the largest eigenvalue of the matrix $\sqrt{P_{ij}P_{ji}}$ \cite{A12b}.
Then, modulo a translation, the current generating function of a dynamics $(p_1,p_2) \in [\alpha]$ is directly given by \cite{A12b}
\bea
q^\alpha = - 2 \log \rho^\alpha  (A) 
\label{q}
\eea
where $\rho^\alpha$ is calculated along the equivalence class $[\alpha]$ (Fig. \ref{fig5}). 
Then, you just need to derive $q^\alpha$ to obtain the current $J^\alpha$ and all higher-order cumulants (you can obtain the exact expression for $q^\alpha$ by inserting Eqs. (\ref{param}) into (\ref{rho})).
While I won't go into details here, it is worth mentioning that the generating function can be used to derive the equivalence classes, instead of the other way around (and this how I realized the existence of these equivalence classes in the first place).

That's all well and good, but what if you're only interested in the fluctuations of a given dynamics $(\bar{p}_1,\bar{p}_2)$, instead of what happens under different nonequilibrium conditions?
It turns out this is actually one and the same thing: 
the generating function of the dynamics $(\bar{p}_1,\bar{p}_2)$ is identical to the generating function (modulo a translation) of any dynamics within its equivalence class $\bar{\alpha}$ \cite{A12b}.
In particular, you can take the equilibrium dynamics $\bar{p}_0 \in \bar{\alpha}$ as your reference dynamics, establishing the equivalence of equilibrium and nonequilibrium fluctuations \cite{A12c}.\\

\section{Neat! What's next and how can I help?}

I hope this provides a more approachable way to understand my recent research and thinking.
I could of course share many additional details but this should be enough to introduce the main concepts. 
Looking forward, I might do a similar exercise for a system with two currents to illustrate the symmetries that arise with coupled currents (see \cite{A22} for a  numerical verification that the response coefficients are fully symmetric). 

From a theoretical point of view, it is plausible that the equivalence classes $[\alpha]$ could be directly derived from the field $\rho$. 
While I haven't explored this in detail, this would provide an interesting (and possibly practical as well) addition to the framework presented here.
Also, the present decomposition and analysis of nonequilibrium dynamics are focused on stationary states. 
Other interesting properties could emerge when looking at time-dependent driving forces within equivalence classes. 
Finally, what is the analogue of equivalence classes for nonequilibrium classical or quantum dynamics?
I would be keen to understand if and how these results extend to other classes of systems.

In any case, happy to collaborate with anyone interested in exploring these issues!

\vskip 0.5 cm

{\bf Disclaimer.} This paper is not intended for journal publication. 

\vskip 0.5 cm

\section*{Appendix}

The space of stochastic dynamics can be partitioned into equivalence classes. 
To this end, we define the operator
\bea
\bar{P}_{ij} = \sqrt{P_{ij} P_{ji}}
\label{PP}
\eea
and the corresponding {\it equivalence relation} 
\bea
P \sim H \quad \text{if there exists a factor} \ \rho \ \text{such that} \quad \bar{P} = \rho  \ \bar{H} \,  .
\label{eqrel}
\eea
An equivalence class $[P] = \{ P' : P' \sim P\}$ is determined by the composition of an equilibrium dynamics $E$ and the set of all possible affinities $\pmb{A}$ (see Ref. \cite{A12b} for a detailed construction).  

For the toy model (\ref{toymodel}), the equivalence classes can be determined explicitly.
F
The simplest approach is to start from an equilibrium dynamics $E = P(\alpha, 1-\alpha)$.
From relation (\ref{eqrel}), any dynamics in the equivalence class $[\alpha]$ satisfies $\sqrt{P_{ij}P_{ji}} = \rho \sqrt{E_{ij}E_{ji}}$.
Taking into account the symmetry of the model, this relation translates into the polynomial system of equations \cite{A23}
\bea
\sqrt{p_1 (1-p_2)} &=& \rho \, \alpha \nonumber \\
\sqrt{p_2 (1-p_1)} &=& \rho \, (1-\alpha) \nonumber
\eea
whose solution corresponds to Eqs (\ref{rho}) and (\ref{alpha}).\\

{\it Note:} In a previous version of this paper, I used a different parametrization of the equivalence classes.
To simplify the logic and increase consistency across papers, I now parametrize the equivalence classes directly in terms of the equilibrium dynamics $(\alpha, 1-\alpha) = (p_0, 1-p_0)$.

\end{document}